\documentclass[12pt]{article}
\usepackage{psfrag}
\usepackage{graphicx}
\usepackage{epstopdf}
\usepackage{amsmath}
\usepackage[T1]{fontenc}
\usepackage[utf8]{inputenc}

\begin{document}

\title{Geometrical description of the phase transition in the Gaździcki Gorenstein model}
\date{\today}
\author{Kacper Zalewski\\Institute of Nuclear Physics, Krakow, \\ Institute of Physics Jagellonian University}
\maketitle
\begin{abstract}
The derivation of the phase transition in the model of Gaździcki and Gorenstein is generalized and simplified by using a geometrical construction.
\end{abstract}

In heavy ion collisions, when the ratio of the number of produced $K^+$ mesons to the number of produced $\pi^+$ mesons is plotted as a function of the cms collision energy per pair of nucleons, a sharp maximum, known as the horn, is observed around $9$ GeV. A compilation of data from the experiments NA49 \cite{ALT}, STAR \cite{KUM} and ALICE \cite{FLO}, showing this effect, can be found in \cite{RUS}.

The horn was predicted by Gaździcki and Gorenstein \cite{GAZ}, who had expected it as a result of  a  transition from the low energy hadronic (white)  W-phase to the high energy Q-phase of deconfined quarks and gluons.  The transition occurs at a temperature of about $200$MeV. In the  W-phase the mass of the strange degrees of freedom is assumed to be $500$MeV, while the other degrees of freedom are massless. This explains the rapid rise of the $K^+/\pi^+$ ratio with increasing temperature, or equivalently energy, below the transition region. In the Q-phase, the ratio of the number of the strange degrees of freedom to the number of the non-strange degrees of freedom is much smaller than in the W-phase. Consequently, the phase transition results in a drop of the $K^+/\pi^+$ ratio and the horn appears. At high temperature, in the Q-phase, the mass of the strange degrees of freedom is smaller than in the W-phase and the temperature is higher. For both these reasons, the increase of the ratio with  energy, is slower than in the W-phase. The point is to describe the phase transition.

Let us note first that the standard Van der Waals -- Maxwell analysis is not applicable here. The equations of state for the two phases can be written in the same form

\begin{equation}\label{}
  (p_i - B_i)V = N_iT, \qquad i = W,Q,
\end{equation}
but the constants $B_i$ are different:

\begin{equation}\label{}
  B_W = 0;\qquad B_Q \approx 0.6\mbox{GeV}.
\end{equation}
In the Van der Waals analysis it is crucial that along an isotherm one can go smoothly, though through unstable states, from the gas phase to the liquid phase. Here, such a transition is not possible, because at  some point the constant $B$ must jump from zero to $0.6$ GeV. Moreover, for the Q and W phases a change of energy determines the corresponding change of volume, while in the Van der Waals gas these two parameters are independent.

In spite of that, Gaździcki and Gorenstein find that the transition is analogous to the familiar  water-vapour transition, as described by the Van der Waals equation with the Maxwell construction. When energy increases, at some point the temperatures and pressures of the two phases become equal and further energy increase is due to changes in the ratio of the volumes occupied by the two phases and not to changes in temperature. At the beginning of this stage, all the volume is occupied by the W-phase and at the end by the Q-phase. From this point on, the Q-phase evolves in the standard way.

The argument given in \cite{GAZ} is rather long and complicated. In the present paper we describe a geometrical demonstration, which makes the result almost obvious. The discussion is more general than the previous analysis. In particular, little is assumed about the properties of the W- and Q- phases.

Let us recapitulate the assumptions of the model which are needed for our analysis, besides the basic assumption \cite{GAZ} that the thermodynamics of equilibrium states is applicable to the early stages of high energy collisions.
\begin{itemize}
  \item For given mass numbers of the colliding nuclei, the energy $E$ and the volume $V$ of the system are determined by the collision energy and do not depend on what phase, or phases, are inside. Consequently, the energy density

      \begin{equation}\label{}
        \epsilon = \frac{E}{V}
      \end{equation}
can be used instead of the collision energy, when studying the energy dependence, and the entropy densities

  \begin{equation}\label{}
    s = \frac{S}{V}
  \end{equation}
instead of the entropies, when comparing the stability of states having the same energy.
      \item All the chemical potentials are zero. Consequently, the free enthalpy density is also zero and for each phase or set of phases:

      \begin{equation}\label{eqpres}
        \epsilon - Ts + p = 0.
      \end{equation}
  \item The system is isolated so that its stable equilibrium corresponds to maximum entropy. Consequently, when the energy is fixed, a state with higher entropy density is stable with respect to any state of lower entropy density.
  \item At low energies $s_W(\epsilon)>s_Q(\epsilon)$ and at high energies $s_W(\epsilon)<s_Q(\epsilon)$. There is exactly one crossing point, where at some $\epsilon=\epsilon^*$

      \begin{equation}\label{}
        s_W(\epsilon^*) = s_Q(\epsilon^*).
      \end{equation}
      \end{itemize}

      For each phase

      \begin{equation}\label{enttem}
        \frac{ds}{d\epsilon} = \frac{1}{T} > 0
      \end{equation}
  and the temperature is an increasing function of energy. Therefore, the entropy densities of the two phases are monotonically increasing, concave upwards functions of the energy density.

  Let us consider a tangent to both curves $s_i(\epsilon)$ touching the curve $s_W(\epsilon)$ at some point $\epsilon_1 < \epsilon^*$ and the curve $s_Q(\epsilon)$ at some point $\epsilon_2>\epsilon^*$. The concavity of the two curves implies that in all the interval

  \begin{equation}\label{rangee}
    \epsilon_1 < \epsilon < \epsilon_2
  \end{equation}
the tangent passes above the $s_W(\epsilon)$ and $s_Q(\epsilon)$ curves.

Consider now the mixed states, where a volume $\lambda V$ i occupied by the Q-phase and a volume $(1-\lambda)V$ by the W-phase, with

\begin{equation}\label{ranlam}
  0 \leq \lambda \leq 1.
\end{equation}
It is assumed that the energy densities in the two phases are respectively $\epsilon_2$ and $\epsilon_1$. Then, the value of the parameter $\lambda$ is related to the energy density by the equation

\begin{equation}\label{}
 \epsilon = \lambda \epsilon_1 + (1-\lambda)\epsilon_2
\end{equation}
 and a solution for $\lambda$ satisfying conditions (\ref{ranlam}) can be found for any energy density from the range (\ref{rangee}). The entropy density of the mixed state is

\begin{equation}\label{}
  s =  \lambda s_Q(\epsilon_1) + (1-\lambda)s_W(\epsilon_2).
\end{equation}
This is the segment of the tangent, in the interval (\ref{rangee}). Since the tangent is a straight line, we have the following identities

\begin{equation}\label{}
 \left(\frac{\partial s_W(\epsilon)}{\partial \epsilon}\right)_{\epsilon=\epsilon_1} = \left(\frac{\partial s_Q(\epsilon)}{\partial \epsilon}\right)_{\epsilon=\epsilon_2} = \frac{s_Q(\epsilon_2)-s_W(\epsilon_1)}{\epsilon_2-\epsilon_1}.
\end{equation}
According to (\ref{enttem}) the first equality means that the temperatures of the two phases are equal and according to (\ref{eqpres}) the second implies that their pressures are equal. Thus, we have shown that in the mixed state the two phases constituting it are in equilibrium and the overall entropy is higher than in either of the single phases. The energy densities $\epsilon_1$ and $\epsilon_2$ can be found by solving the equation system

\begin{equation}\label{}
  T_W(\epsilon_1) =T_Q(\epsilon_2),\qquad p_W(\epsilon_1) = p_Q(\epsilon_2).
\end{equation}

Note that we have only proved that in the energy density range (\ref{rangee}) the mixed states are stable with respect to the phases W and Q at the same energy. The argument does not exclude that another phase, e.g. a phase where the coloured and white degrees of freedom coexist, is even more stable.

\textbf{\emph{Acknowledgement}}

The author was partly supported by the Polish National Science Center (NCN) under grant DEC-2013/09/B/ST2/00497.

\end{document}